\documentclass[doublecol]{epl2}
\usepackage{amssymb,amsmath,graphicx,setspace,color,rotating,subfigure,url}
\title{Effects of long memory in the order submission process on the properties of recurrence intervals of large price fluctuations}
\shorttitle{Effects of long memory on recurrence interval properties}

\author{Hao Meng\inst{1,2,3} \and Fei Ren\inst{1,2,3} \and Gao-Feng Gu\inst{1,3} \and Xiong Xiong\inst{4} \and Yong-Jie Zhang\inst{4} \and Wei-Xing Zhou\inst{1,2,3}\footnote{e-mail: wxzhou@ecust.edu.cn} \and Wei Zhang\inst{4}\footnote{e-mail: weiz@tju.edu.cn}}
\shortauthor{H. Meng \etal}

\institute{
  \inst{1} School of Business, East China University of Science and Technology, Shanghai 200237, China\\
  \inst{2} School of Science, East China University of Science and Technology, Shanghai 200237, China\\
  \inst{3} Research Center for Econophysics, East China University of Science and Technology, Shanghai 200237, China\\
  \inst{4} School of Management, Tianjin University, Tianjin 300072, China
}


\abstract{Understanding the statistical properties of recurrence intervals of extreme events is crucial to risk assessment and management of complex systems. The probability distributions and correlations of recurrence intervals for many systems have been extensively investigated. However, the impacts of microscopic rules of a complex system on the macroscopic properties of its recurrence intervals are less studied. In this Letter, we adopt an order-driven stock market model to address this issue for stock returns. We find that the distributions of the scaled recurrence intervals of simulated returns have a power law scaling with stretched exponential cutoff and the intervals possess multifractal nature, which are consistent with empirical results. We further investigate the effects of long memory in the directions (or signs) and relative prices of the order flow on the characteristic quantities of these properties. It is found that the long memory in the order directions (Hurst index $H_s$) has a negligible effect on the interval distributions and the multifractal nature. In contrast, the power-law exponent of the interval distribution increases linearly with respect to the Hurst index $H_x$ of the relative prices, and the singularity width of the multifractal nature fluctuates around a constant value when $H_x<0.7$ and then increases with $H_x$. No evident effects of $H_s$ and $H_x$ are found on the long memory of the recurrence intervals. Our results indicate that the nontrivial properties of the recurrence intervals of returns are mainly caused by traders' behaviors of persistently placing new orders around the best bid and ask prices.
}

\begin{document}

\maketitle

\section{\label{S1:Intro}Introduction}

Most natural and socioeconomic systems are complex systems in which extreme events occur more frequently than Gaussian and exhibit complex behaviors \cite{Bunde-Kropp-Schellnhuber-2002,Sornette-2003,Sornette-2004}. Such extreme events usually cause tremendous damages on the systems, incurring losses of lives, belongings and properties. Hence, understanding the behaviors of extreme events is crucial to the risk assessment and management of complex systems. However, it is not an easy task due to the lack of statistics of extreme events. One strategy to overcome this difficulty is to investigate the evolution of the statistical properties of recurrence intervals of less extreme events and extreme events and make inference by extrapolations \cite{Bunde-Eichner-Havlin-Kantelhardt-2003-PA,Bunde-Eichner-Havlin-Kantelhardt-2004-PA,Bunde-Eichner-Kantelhardt-Havlin-2005-PRL,Yamasaki-Muchnik-Havlin-Bunde-Stanley-2005-PNAS}.

The recurrence interval analysis has been performed on different time series recorded in diverse systems, including simulated time series with long-term correlations \cite{Bunde-Eichner-Havlin-Kantelhardt-2003-PA,Altmann-Kantz-2005-PRE,Pennetta-2006-EPJB,Olla-2007-PRE,Eichner-Kantelhardt-Bunde-Havlin-2007-PRE,Bogachev-Eichner-Bunde-2007-PRL,Bogachev-Eichner-Bunde-PAG-2008,Santhanam-Kantz-2008-PRE,Moloney-Davidsen-2009-PRE}, climate records \cite{Bunde-Eichner-Havlin-Kantelhardt-2004-PA,Bunde-Eichner-Kantelhardt-Havlin-2005-PRL}, earthquakes \cite{Livina-Tuzov-Havlin-Bunde-2005-PA,Livina-Havlin-Bunde-2005-PRL,Lennartz-Livina-Bunde-Havlin-2008-EPL}, heartbeats \cite{Bogachev-Kireenkov-Nifontov-Bunde-2009-NJP}, energy dissipation rate in three-dimensional fully developed turbulence \cite{Liu-Jiang-Ren-Zhou-2009-PRE}, and volatilities \cite{Kaizoji-Kaizoji-2004a-PA,Yamasaki-Muchnik-Havlin-Bunde-Stanley-2005-PNAS,Lee-Lee-Rikvold-2006-JKPS,Wang-Yamasaki-Havlin-Stanley-2006-PRE,Wang-Weber-Yamasaki-Havlin-Stanley-2007-EPJB,VodenskaChitkushev-Wang-Weber-Yamasaki-Havlin-Stanley-2008-EPJB,Ren-Zhou-2008-EPL,Jung-Wang-Havlin-Kaizoji-Moon-Stanley-2008-EPJB,Qiu-Guo-Chen-2008-PA,Wang-Yamasaki-Havlin-Stanley-2008-PRE,Ren-Guo-Zhou-2009-PA,Ren-Gu-Zhou-2009-PA,Wang-Yamasaki-Havlin-Stanley-2009-PRE,Zhang-Wang-Shao-2010-ACS,Joen-Moon-Oh-Yang-Jung-2010-JKPS}, returns \cite{Yamasaki-Muchnik-Havlin-Bunde-Stanley-2006-inPFE,Bogachev-Eichner-Bunde-2007-PRL,Bogachev-Bunde-2008-PRE,Muchnik-Bunde-Havlin-2009-PA,Ren-Zhou-2010-NJP,Ludescher-Tsallis-Bunde-2011-EPL,He-Chen-2011b-PA}, and trading volumes \cite{Podobnik-Horvatic-Petersen-Stanley-2009-PNAS,Ren-Zhou-2010-PRE,Li-Wang-Havlin-Stanley-2011-PRE} in financial markets.
For financial returns, the recurrence intervals are distributed in power-law forms and possess long-term memories \cite{Yamasaki-Muchnik-Havlin-Bunde-Stanley-2006-inPFE,Bogachev-Eichner-Bunde-2007-PRL,Bogachev-Bunde-2008-PRE,Muchnik-Bunde-Havlin-2009-PA,Ren-Zhou-2010-NJP,Ludescher-Tsallis-Bunde-2011-EPL}. These empirical results are crucial to the risk assessment of large price fluctuations \cite{Yamasaki-Muchnik-Havlin-Bunde-Stanley-2006-inPFE,Bogachev-Eichner-Bunde-2007-PRL,Bogachev-Bunde-2009-PRE,Ren-Zhou-2010-NJP,Ludescher-Tsallis-Bunde-2011-EPL}.

The effects of linear and nonlinear long-term correlations of the original time series on the statistical properties of the recurrence intervals have been investigated, which is at the macroscopic level. It is found that the linear long-term correlations in the original time series cause the long memory in the recurrence intervals \cite{Yamasaki-Muchnik-Havlin-Bunde-Stanley-2005-PNAS}. If the original records contain multifractal nature, the distribution of the recurrence intervals decays as a power law \cite{Bogachev-Eichner-Bunde-2007-PRL}; Otherwise, the distribution has a Weibull form. However, an investigation of the microscopic factors that lead to those statistical properties of recurrence intervals still lacks. This Letter contributes to this topic to understand the effects of long memory in the order submission process on the properties of recurrence intervals of stock returns using a microscopic stock trading model.

The price return is defined as the logarithmic difference of stock prices. We normalize the return records by dividing its standard deviation
to obtain the normalized return time series $R(t)$. The recurrence interval $r_{Q}$ is the waiting time between two consecutive normalized price returns above a positive threshold $Q>0$. Since the distributions of recurrence intervals for positive and negative thresholds are symmetric, we present the results for the return intervals between large positive price returns.

\section{Modified Mike-Farmer model}
\label{S1:MMF}

There are various microscopic models for stock price formation processes, most of which are agent-based models constructed according to intuitive and/or empirical regularities and designed for quote-driven market \cite{Zhou-2007}. There are also order-driven market models \cite{Maslov-2000-PA,Preis-Golke-Paul-Schneider-2006-EPL,Preis-Virnau-Paul-Schneider-2009-NJP,Bartolozzi-2010-EPJB}, and to our knowledge, the empirical behavioral model proposed by Mike and Farmer is the first and most realistic model \cite{Mike-Farmer-2008-JEDC}. The Mike-Farmer model mimics the processes of order placement and order cancelation, and orders are executed according to the price-time priority rule. The processes of order placement and cancelation are simulated based on the statistical properties extracted from real order flows of stocks. To submit an order, one needs to determine its direction (buy or sell), price and size. To cancel an order on the limit order book, one needs to identify possible factors causing a trader to withdraw his/her submitted orders. The framework of the Mike-Farmer model opens a promising direction for computational experimental fiance since it allows us to understand the macroscopic properties of stocks from the realistic microscopic level.

The Mike-Farmer model integrates two statistical regularities \cite{Mike-Farmer-2008-JEDC}, {\it{i.e.}}, the long memory of order directions characterized by its Hurst index $H_s$ and the Student distribution of the relative logarithmic order price $x$, which is defined as the logarithmic difference of the order price to the same best price. Let $\pi(t)$ be the logarithmic price of a submitted order at time $t$, and $\pi_{b}(t)$ and $\pi_{a}(t)$ be the logarithms of the current best bid price and best ask price. The relative price $x(t)$ at time $t$ can be defined as $x(t)=\pi(t)-\pi_{b}(t)$ for buy orders and $x(t)=\pi_{a}(t)-\pi(t)$ for sell orders. This model is able to reproduce excellent return distributions at different time scales comparable to real data and the absence of long memory in the return time series \cite{Mike-Farmer-2008-JEDC,Gu-Zhou-2009-EPJB}.

However, the Mike-Farmer model fails to reproduce the long-memory property of volatility \cite{Mike-Farmer-2008-JEDC}. After further including in the model another stylized fact that the relative prices of submitted orders are long-term correlated, Gu and Zhou proposed a modified Mike-Farmer model which is able to capture nicely the long memory in the volatility without distorting the power-law tails in the return distributions and the absence of memory in the return series \cite{Gu-Zhou-2009-EPL}. The return time series generated from the modified Mike-Farmer model also exhibit the well-known multifractal nature \cite{Gu-Zhou-2009-EPL}. It is also found that a Poisson cancelation process is sufficient to obtain these stylized facts \cite{Gu-Zhou-2009-EPJB,Gu-Zhou-2009-EPL}. In this Letter, we adopt this modified Mike-Farmer model. Specifically, the order directions are simulated from the signs of the increments of a fractional Brownian motion of Hurst index $H_s$. The relative order prices are drawn from a Student distribution and then manipulated to introduce long-term correlations of Hurst index $H_x$ using the amplitude adjusted Fourier transform algorithm \cite{Schreiber-Schmitz-1996-PRL}. The order size is fixed to be unity \cite{Mike-Farmer-2008-JEDC,Gu-Zhou-2009-EPJB,Gu-Zhou-2009-EPL}.

\begin{figure}[tb]
\centering
\includegraphics[width=4cm]{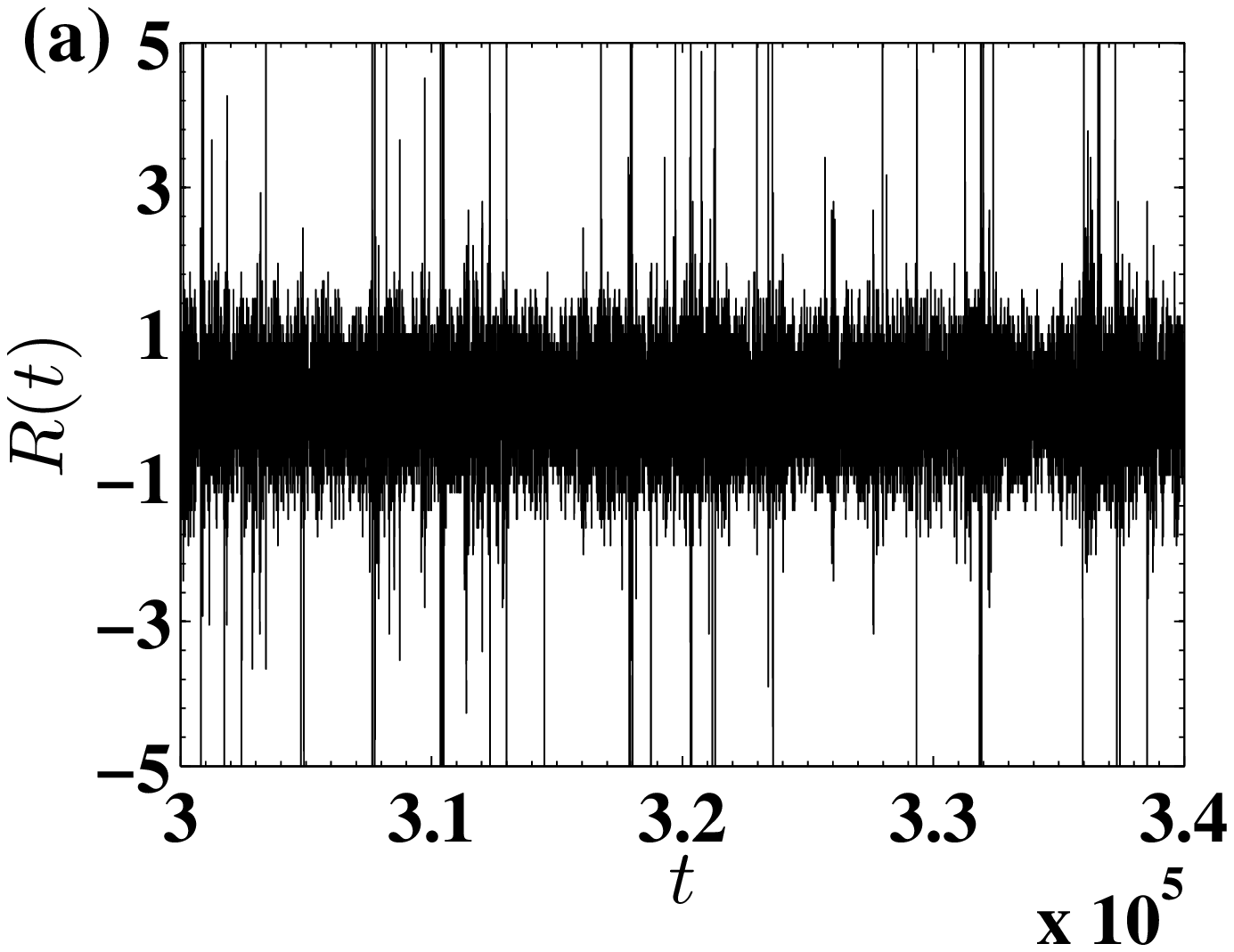}
\includegraphics[width=4cm]{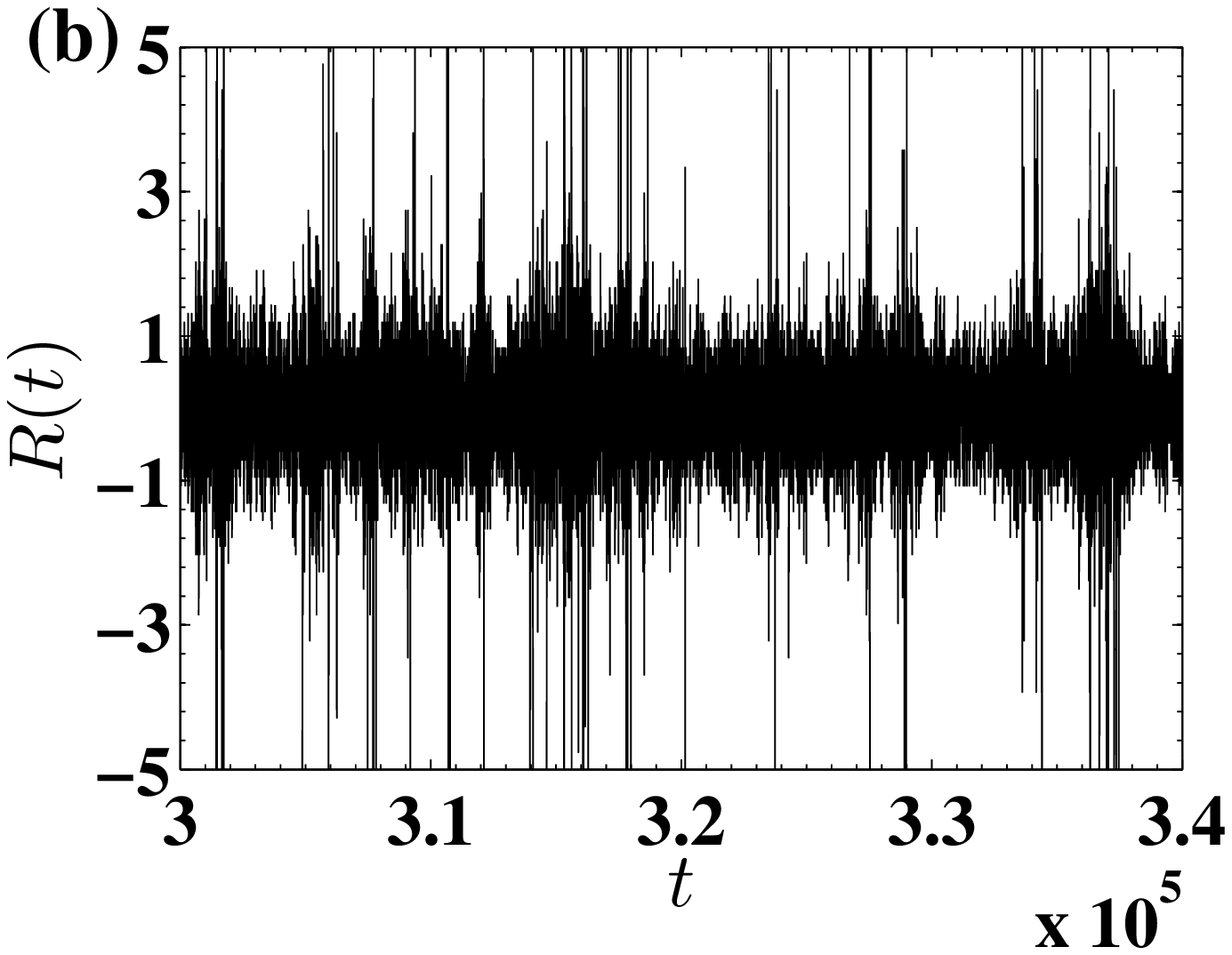}\\
\includegraphics[width=4cm]{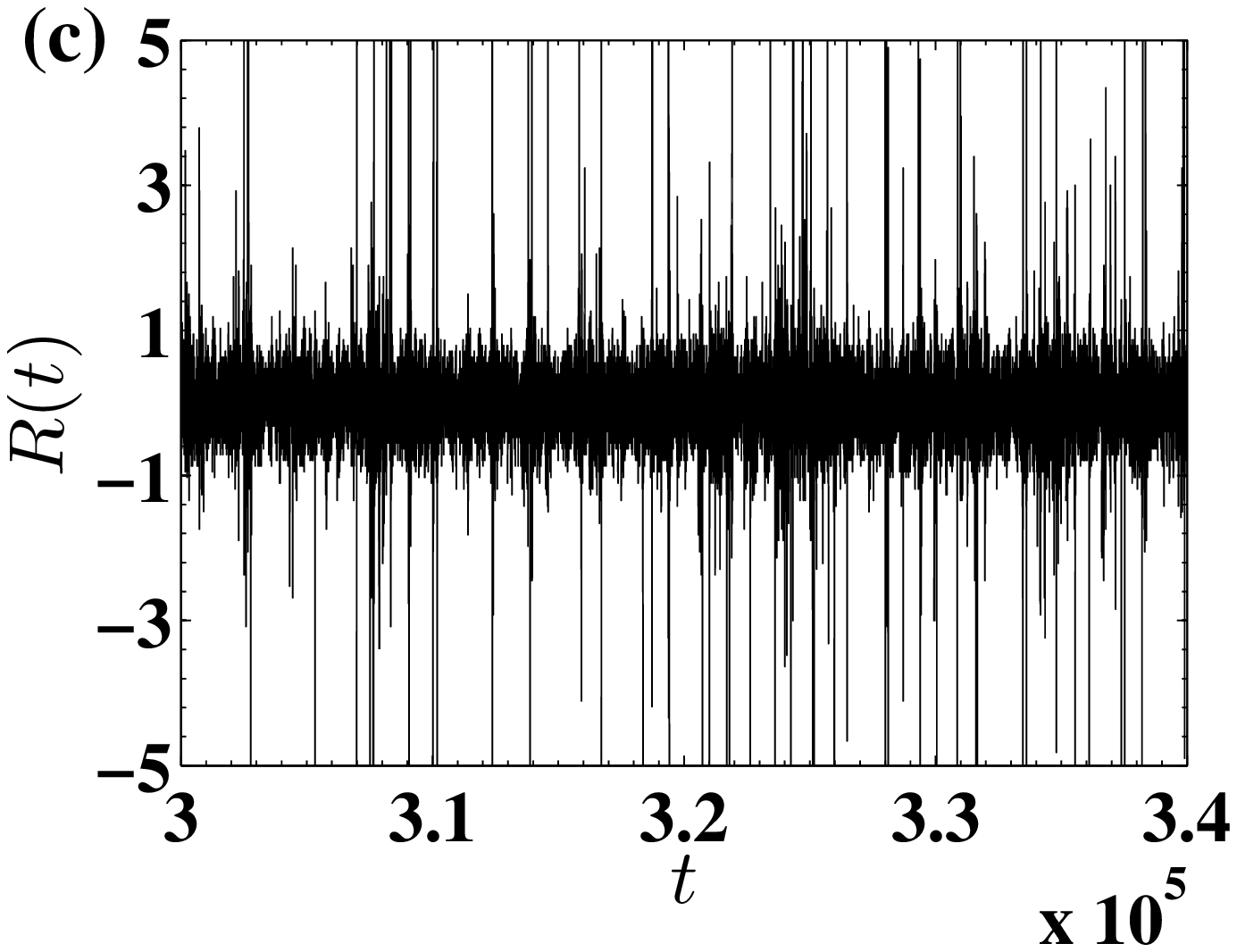}
\includegraphics[width=4cm]{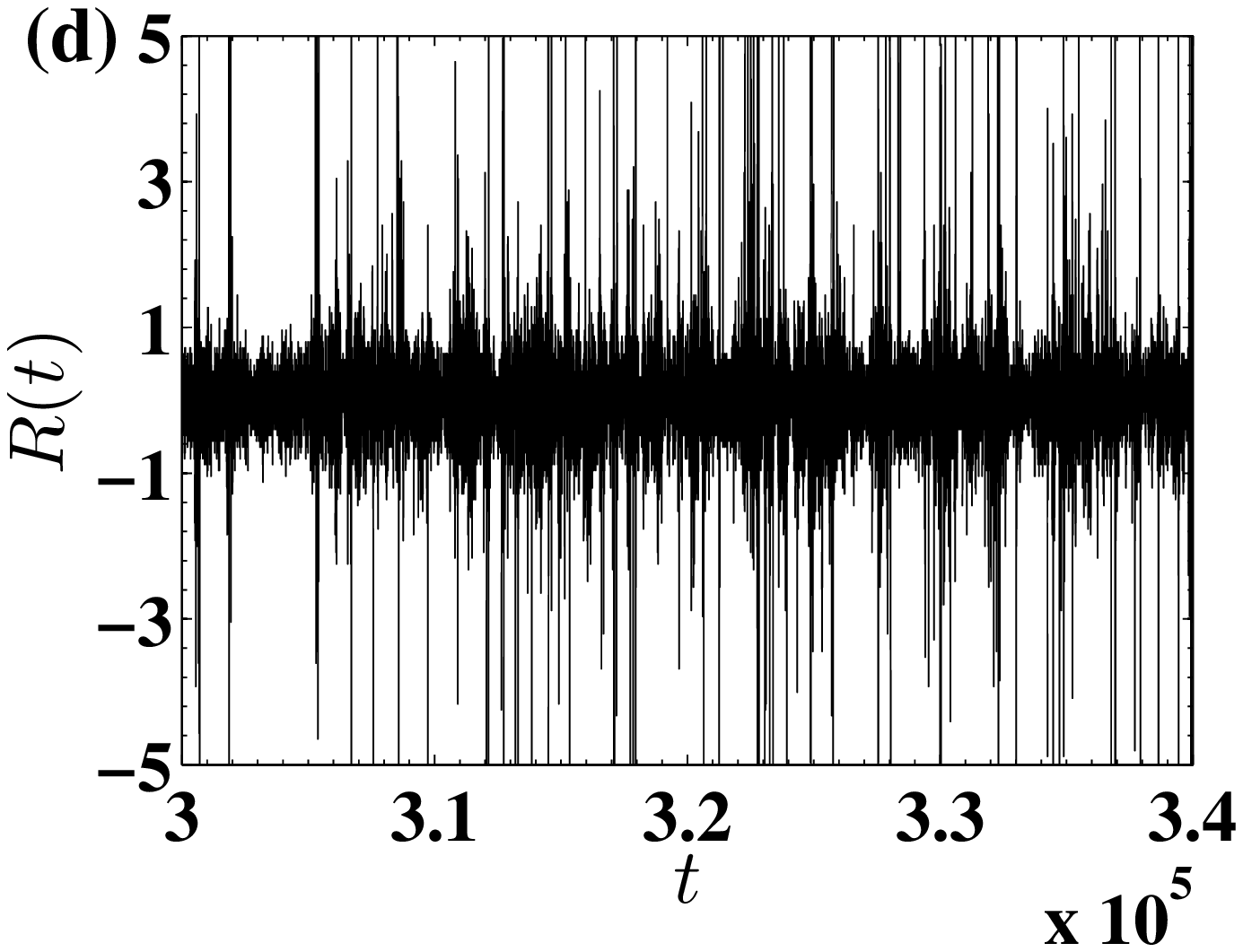}
\caption{\label{Fig:ReturnSeries} Realization segments of standardized returns $R(t)$ simulated from the modified Mike-Farmer model for different values of $H_s$ and $H_x$: (a) $H_x=0.5$ and $H_s=0.5$, (b)$H_x=0.75$ and $H_s=0.5$, (c) $H_x=0.5$ and $H_s=0.75$, and (d) $H_x=0.75$ and $H_s=0.75$.}
\end{figure}

In our simulations, the distribution of the relative prices is fixed to be a symmetric Student distribution with the freedom degree being 1.3 and the scale parameter 0.0024 \cite{Mike-Farmer-2008-JEDC}. Actually, our numerical experiments indicate that these parameters have negligible impacts on the statistical properties of the recurrence intervals. We vary the values of $H_s$ and $H_x$ from 0.5 to 0.9. For each round of simulation, there are $2\times10^{6}$ steps, and about 500,000 returns are recorded for analysis. For each pair of $H_s$ and $H_x$, five rounds of simulations are performed and the results presented below are averaged with these repeated simulations. Figure \ref{Fig:ReturnSeries} illustrates four segments of return realizations for different values of $H_s$ and $H_x$, where the original returns are standardized.

\section{Scaled PDF of recurrence intervals}
\label{S1:PDF}

\begin{figure}[tb]
\centering
\includegraphics[width=4cm]{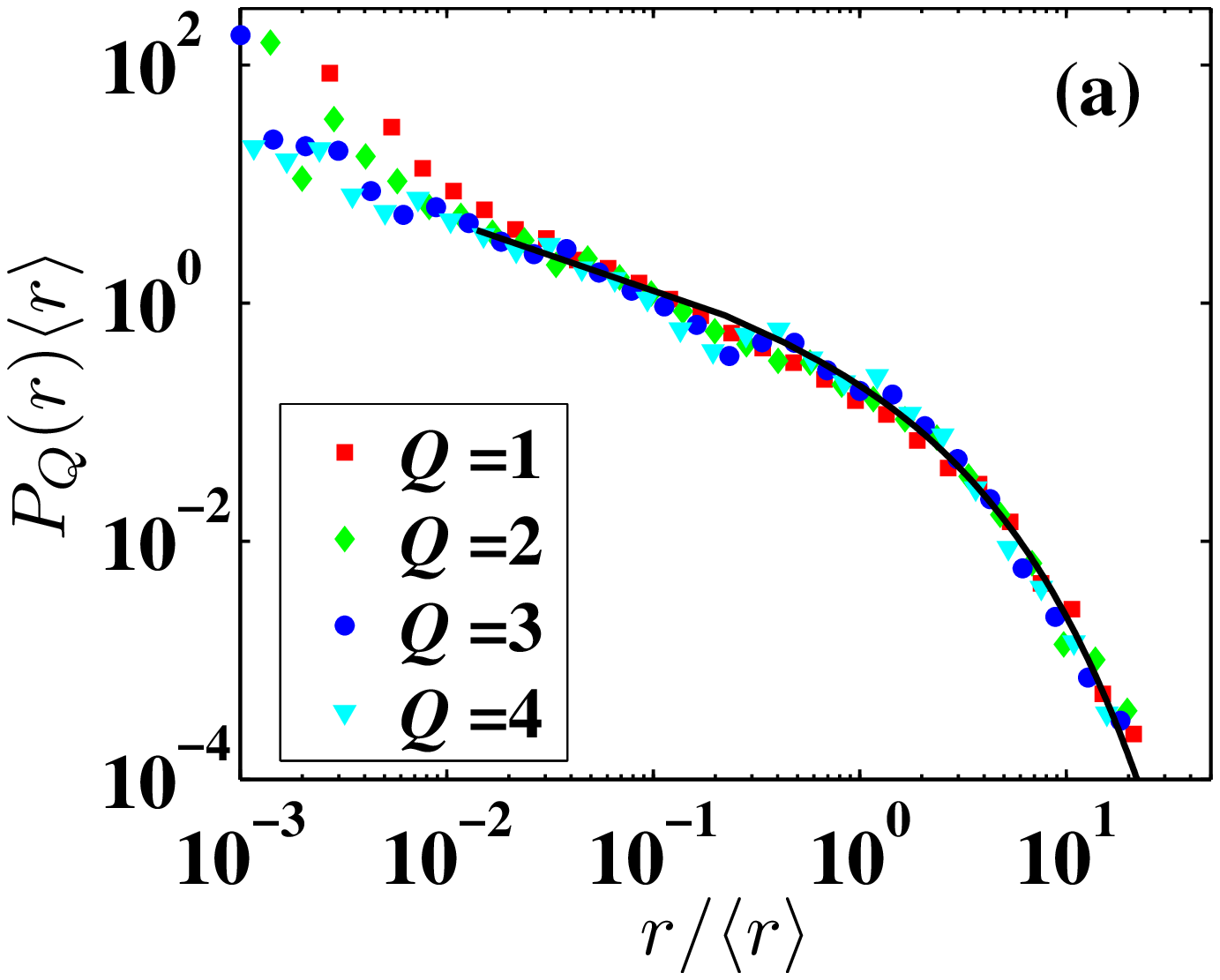}
\includegraphics[width=4cm]{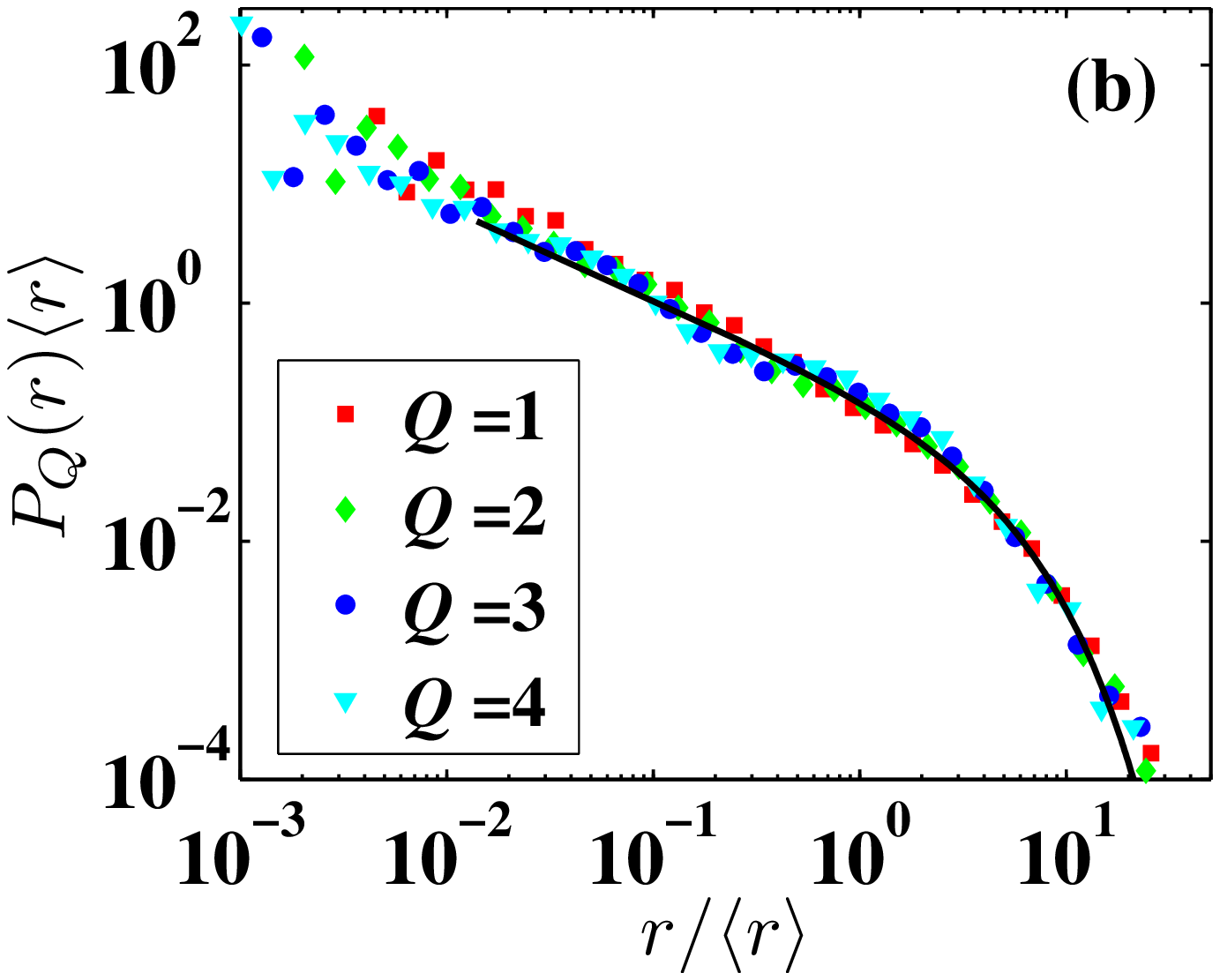}
\includegraphics[width=4cm]{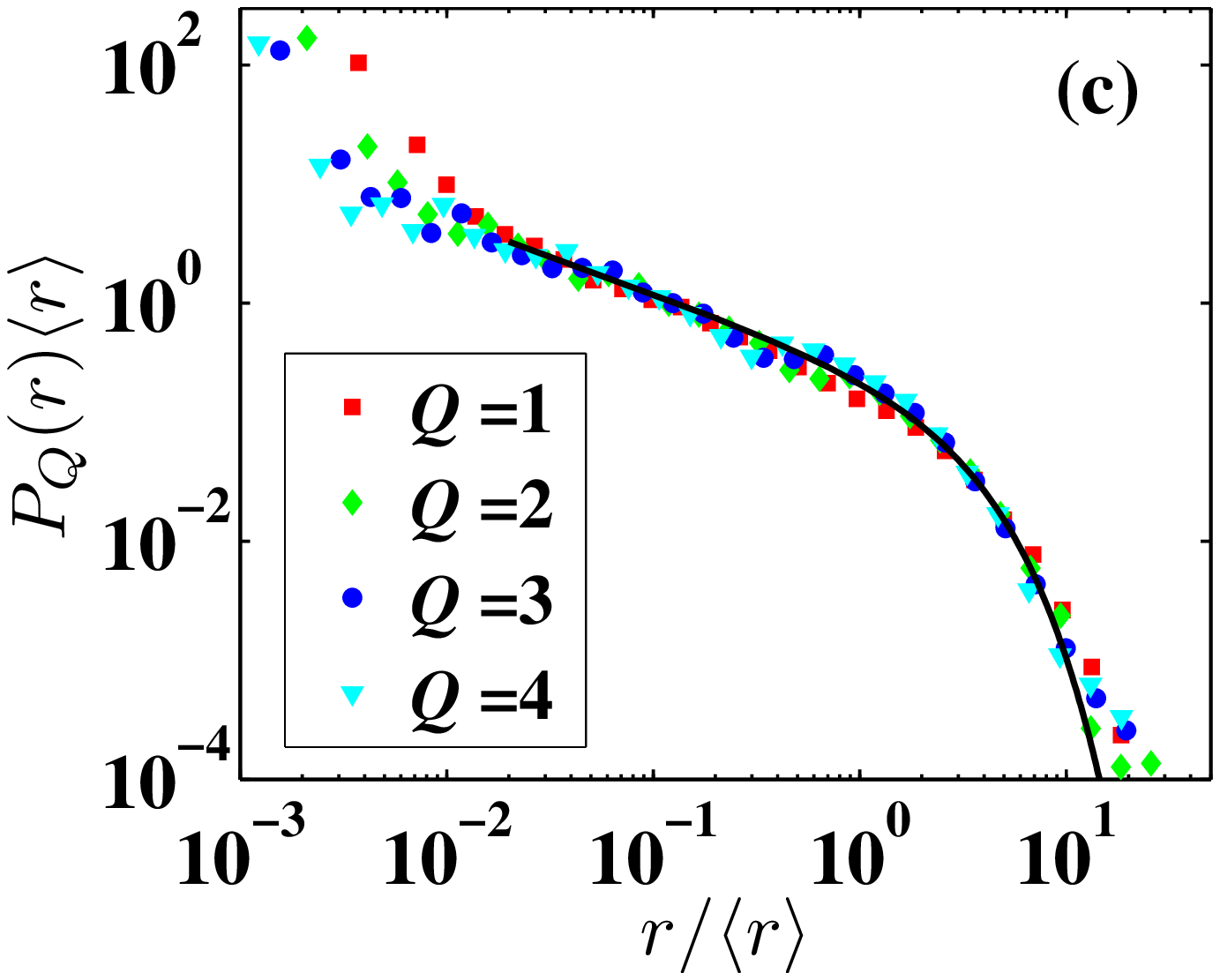}
\includegraphics[width=4cm]{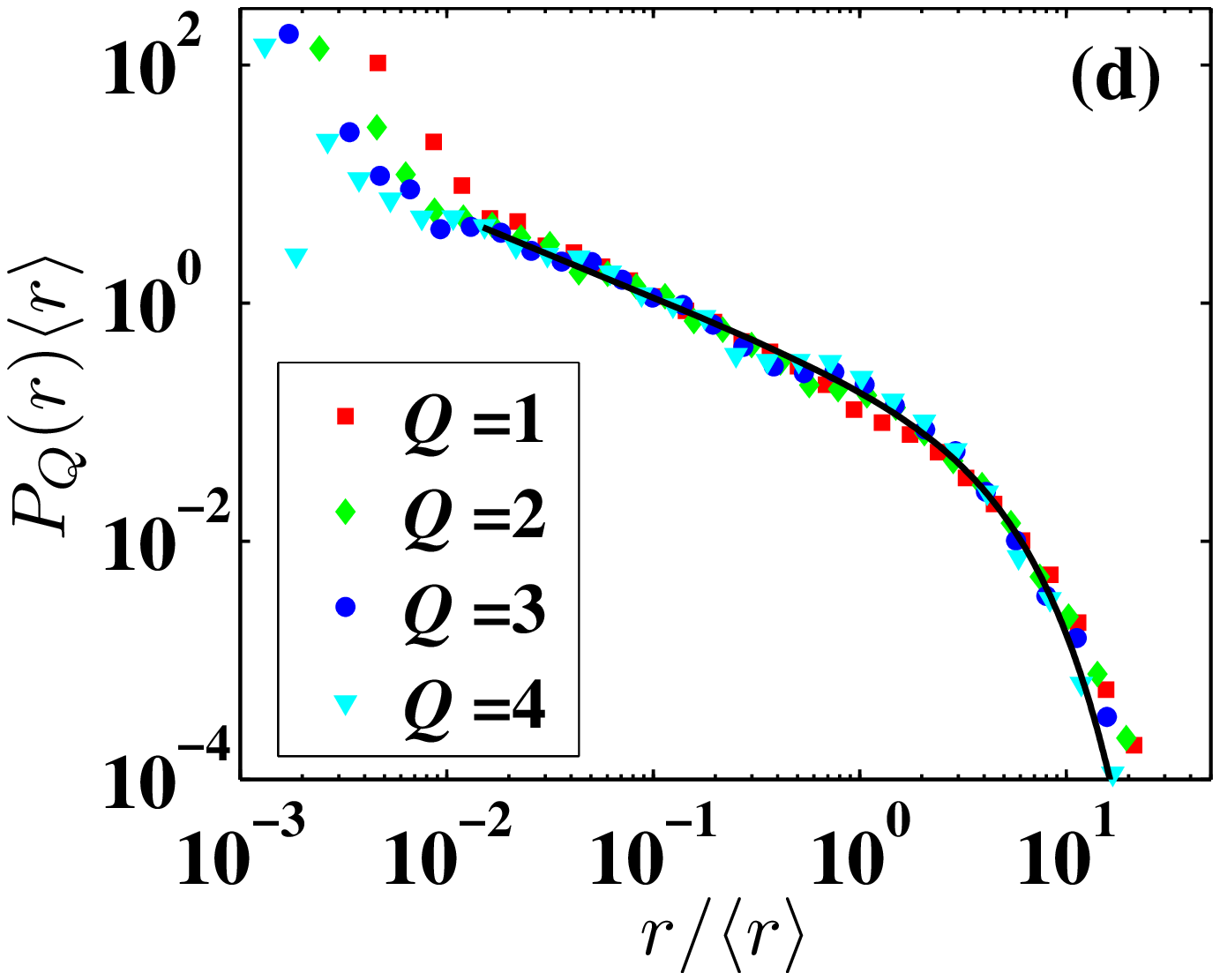}
\caption{\label{Fig:4PDFs} (Color online) Scaled probability distributions of the return recurrence intervals above positive thresholds $Q=1,2,3$ and $4$ for four pairs of $H_s$ and $H_x$: (a) $H_x=0.5$ and $H_s=0.5$, (b)$H_x=0.75$ and $H_s=0.5$, (c) $H_x=0.5$ and $H_s=0.75$, and (d) $H_x=0.75$ and $H_s=0.75$. The solid lines are the fits to the generalized Gamma function.}
\end{figure}

We first investigate the probability distribution functions (PDFs) $P_{Q}(r)$ of return intervals. For each group of model parameters, we perform five rounds of simulations and calculate five time series of recurrence intervals for different thresholds $Q$. For each $Q$, the interval time series are divided by their corresponding mean values $\langle r \rangle$ and put them together. We find that there is no asymmetry between the two recurrence interval PDFs for $Q$ and $-Q$, which is consistent with empirical data \cite{Ren-Zhou-2010-NJP}. We thus present the results for positive thresholds $Q>0$. Figure \ref{Fig:4PDFs} illustrates the scaled PDFs $\langle r \rangle P_Q(r)$ as a function of scaled return intervals $r/\langle r \rangle$ above four thresholds $Q=2$, $3$, $4$ and $5$ for four pairs of $H_x$ and $H_s$.

As we can see in each plot of Fig. \ref{Fig:4PDFs}, the scaled PDFs for different thresholds $Q$ differ from each other for small $r/\langle r \rangle$. When $r/\langle r \rangle>10^{-2}$, the PDFs in each plot collapse onto a single master curve for different thresholds. In other words, we have
\begin{equation}
  \langle r \rangle P_{Q}(r) = p(r/\langle r \rangle),
\end{equation}
where $p(x)$ is a master curve independent of $Q$ for large $r/\langle r \rangle$. We find that $p(x)$ can be well fitted using a generalized Gamma function
\begin{equation}
  p(x) = Ax^{-\beta}e^{-\gamma x^{\delta}},
  \label{Eq:px}
\end{equation}
where $\beta$, $\gamma$ and $\delta$ are parameters and $A$ is the normalization factor
\begin{equation}
  A = \frac{\delta}{\gamma^{\frac{\beta-1}{\delta}}\Gamma(\frac{1-\beta}{\delta})}.
\end{equation}
We fitted the generalized Gamma distribution for different values of $H_x$ and $H_s$. There is no clear evidence that $\gamma$ and $\delta$ depends on $H_x$ or $H_s$. In contrast, Table \ref{Table:beta} shows that the power-law exponent $\beta$ increases with $H_x$ for each $H_s$. For fixed $H_x$, the relation between $\beta$ and $H_s$ is quite complicated. When the relative prices are uncorrelated ($H_x\approx0.5$), the exponent $\beta$ increases with $H_s$. When the relative prices are positively correlated ($H_x>0.5$), $\beta$ shows an inverse V-shape with respect to $H_s$.

\begin{table}
  \centering
  \caption{\label{Table:beta} Fitted power-law exponents $\beta$ in Eq. (\ref{Eq:px}) for different values of $H_x$ and $H_s$.}
  \medskip
  \begin{tabular}{cccccccccc}
    \hline\hline
  && \multicolumn{5}{c}{$H_s$}& \\
  \cline{3-7}
      $H_x$ && $0.50$  & $0.60$  & $0.70$  & $0.80$  & $0.90$  \\
    \hline
    $0.50$  && $0.466$  & $0.509$  & $0.619$ & $0.629$ & $0.610$ \\
    $0.60$  && $0.567$  & $0.646$  & $0.638$ & $0.643$ & $0.606$ \\
    $0.70$  && $0.694$  & $0.717$  & $0.730$ & $0.643$ & $0.677$ \\
    $0.80$  && $0.774$  & $0.803$  & $0.773$ & $0.759$ & $0.744$ \\
    $0.90$  && $0.835$  & $0.868$  & $0.858$ & $0.823$ & $0.785$ \\
    \hline\hline
  \end{tabular}
\end{table}

The dependence of $\beta$ on $H_x$ can be explained qualitatively. Numerical simulations show that the Hurst index of the absolute returns increases with $H_x$ \cite{Gu-Zhou-2009-EPL}. Hence, the clustering phenomenon becomes more significant for large $H_x$, as illustrated in Fig.~\ref{Fig:ReturnSeries}. Fixing the threshold $Q$, there are more small recurrence intervals and less large recurrence intervals for large $H_x$. Therefore, the distribution of the recurrence intervals decays faster for larger $H_x$, resulting in a larger power-law exponent $\beta$. On the contrary, there is no clear dependence between the Hurst index of absolute returns and the memory effect of order directions \cite{Gu-Zhou-2009-EPL}. This might explain the ambiguous relationship between $\beta$ and $H_s$.

\begin{figure}[b!]
  \centering
  \includegraphics[width=7cm]{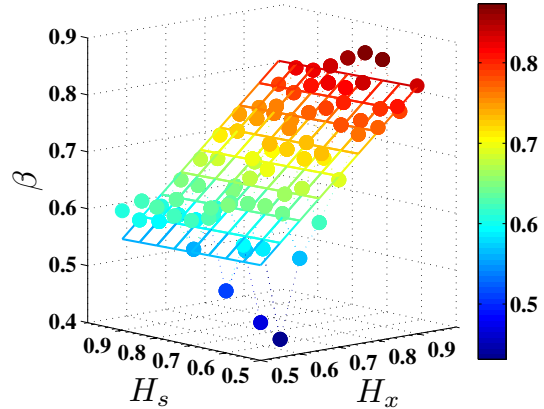}
  \caption{\label{Fig:beta:Hs:Hx} (Color online) Dependence of the power-law exponent $\beta$ on $H_x$ and $H_s$. The planar grid is the best fit of the data points to $\beta = a + bH_x + cH_s$, where $a=0.243$, $b=0.665$ and $c=-0.0174$.}
\end{figure}

If we plot $\beta$ against $H_x$ for fixed $H_s$, we observe nice linear relationship, as shown in Fig.~\ref{Fig:beta:Hs:Hx}. We fit the data points in Fig.~\ref{Fig:beta:Hs:Hx} to the following linear equation
\begin{equation}
  \beta = a + bH_x + cH_s.
\end{equation}
We find that this equation can be accepted at the significance level of $5\%$. The fitted coefficients are $a=0.243$, $b=0.665$ and $c=-0.0174$. A $t$-test for the coefficients $a$, $b$ and $c$ gives that the $p$-values are $p_{a}=p_{b}=0.00$ and $p_{c}=0.60$. It means that the coefficient $b$ is significantly different from zero while $c$ is insignificant. These results are consistent with the observations from table \ref{Table:beta}.


\section{Long-term correlation in recurrence intervals}
\label{S1:Correlation}

There is evidence showing that the recurrence intervals of positive and negative stock returns exhibit long-term correlations \cite{Ren-Zhou-2010-NJP}. Although the return time series is uncorrelated, the time series of all positive or negative returns are correlated. This long-term correlation is the source of the long-term correlation in the recurrence intervals of returns \cite{Ren-Zhou-2010-NJP}. This is similar as for financial volatility, where the long-term correlation in the intervals stems from the long-term correlation in the volatility series \cite{Yamasaki-Muchnik-Havlin-Bunde-Stanley-2005-PNAS}.

\begin{figure}[t!]
\centering
\includegraphics[width=4cm]{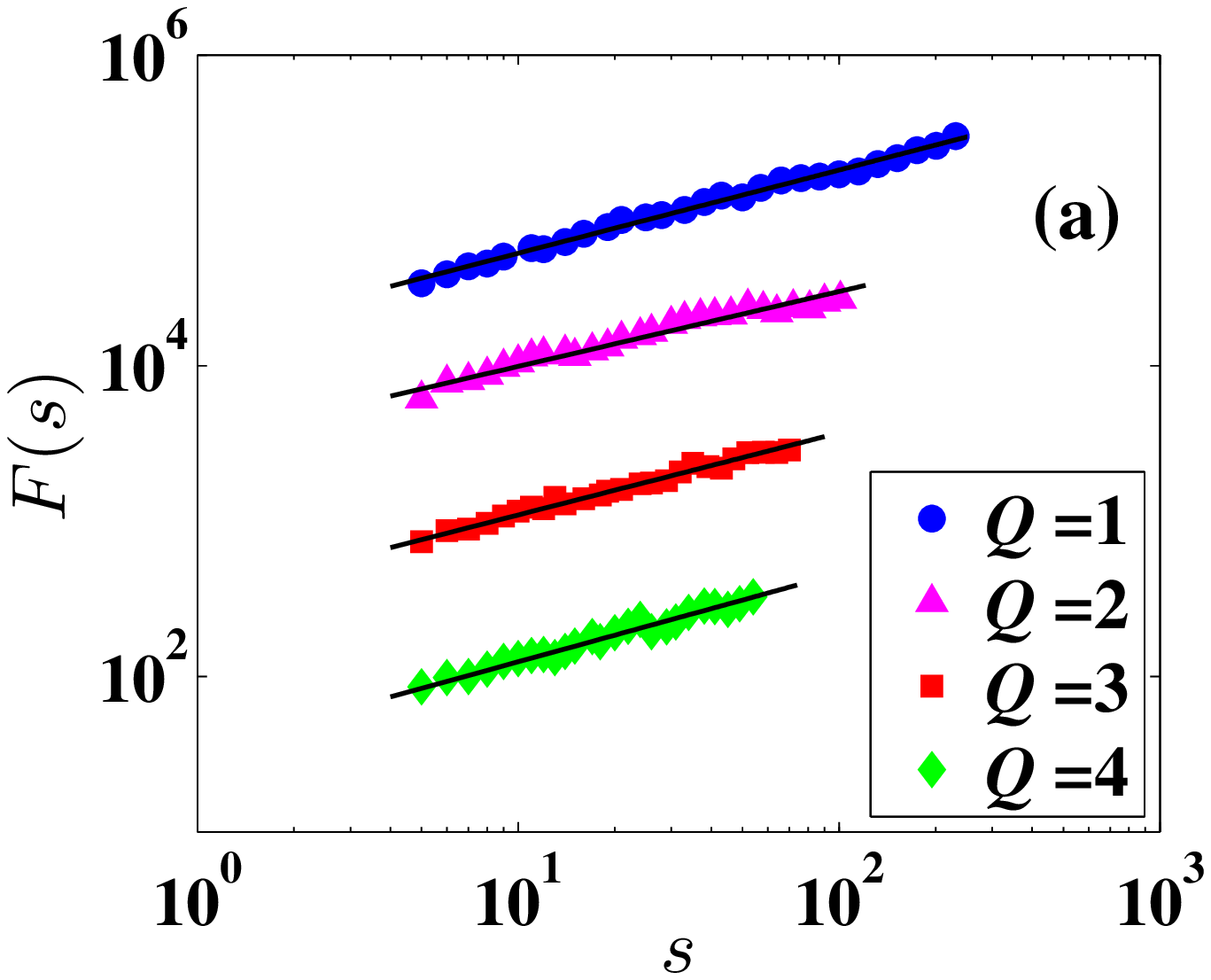}
\includegraphics[width=4cm]{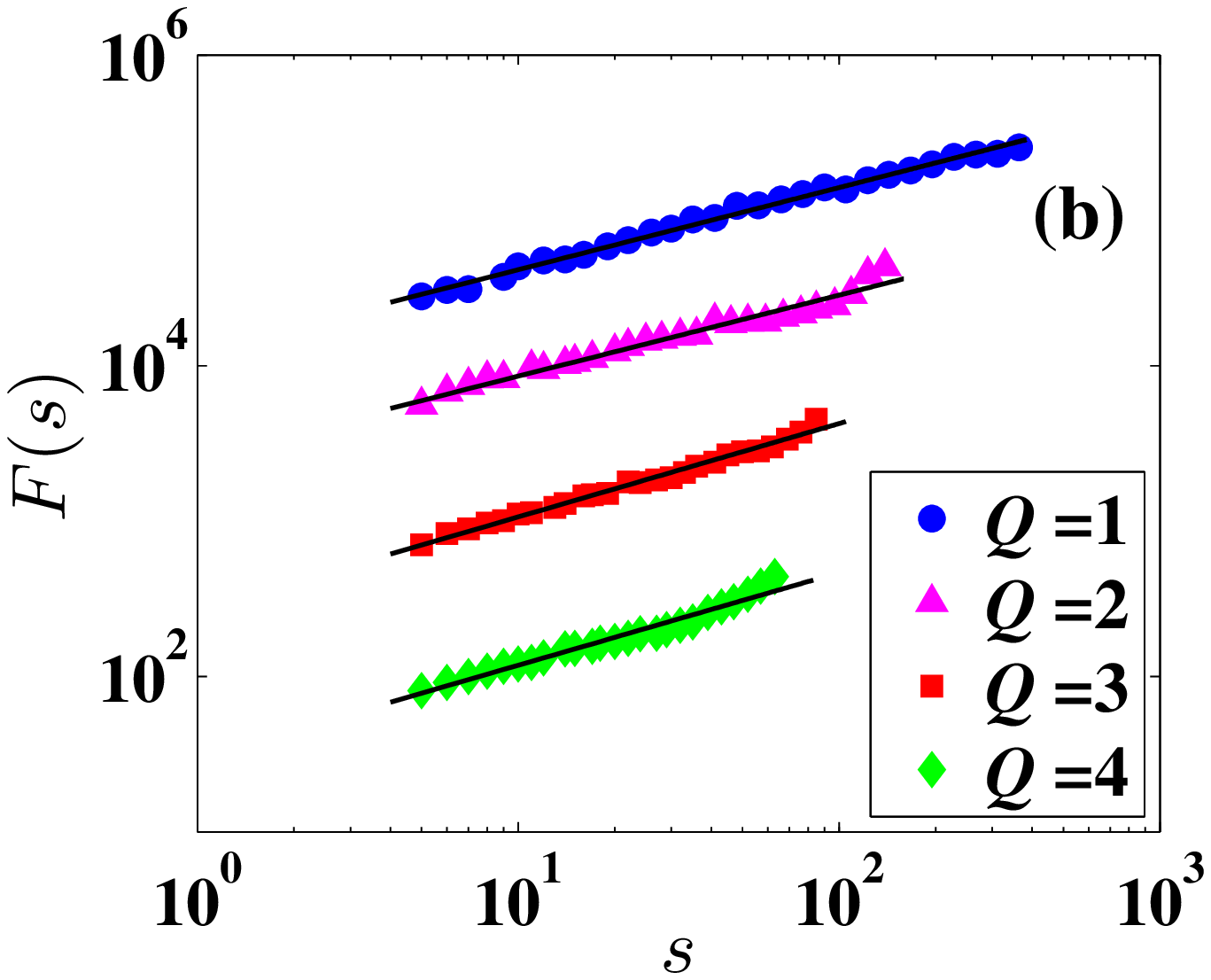}
\includegraphics[width=4cm]{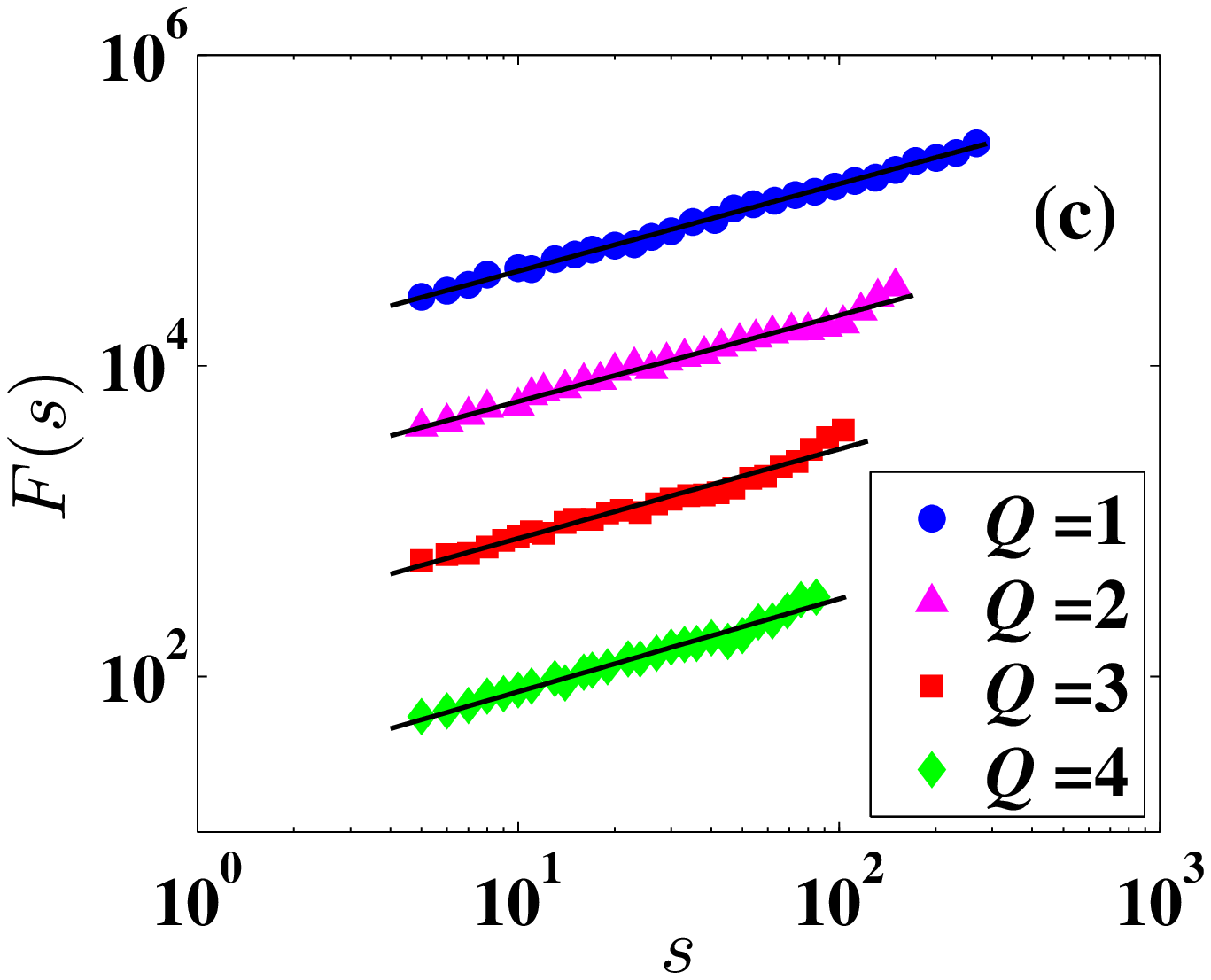}
\includegraphics[width=4cm]{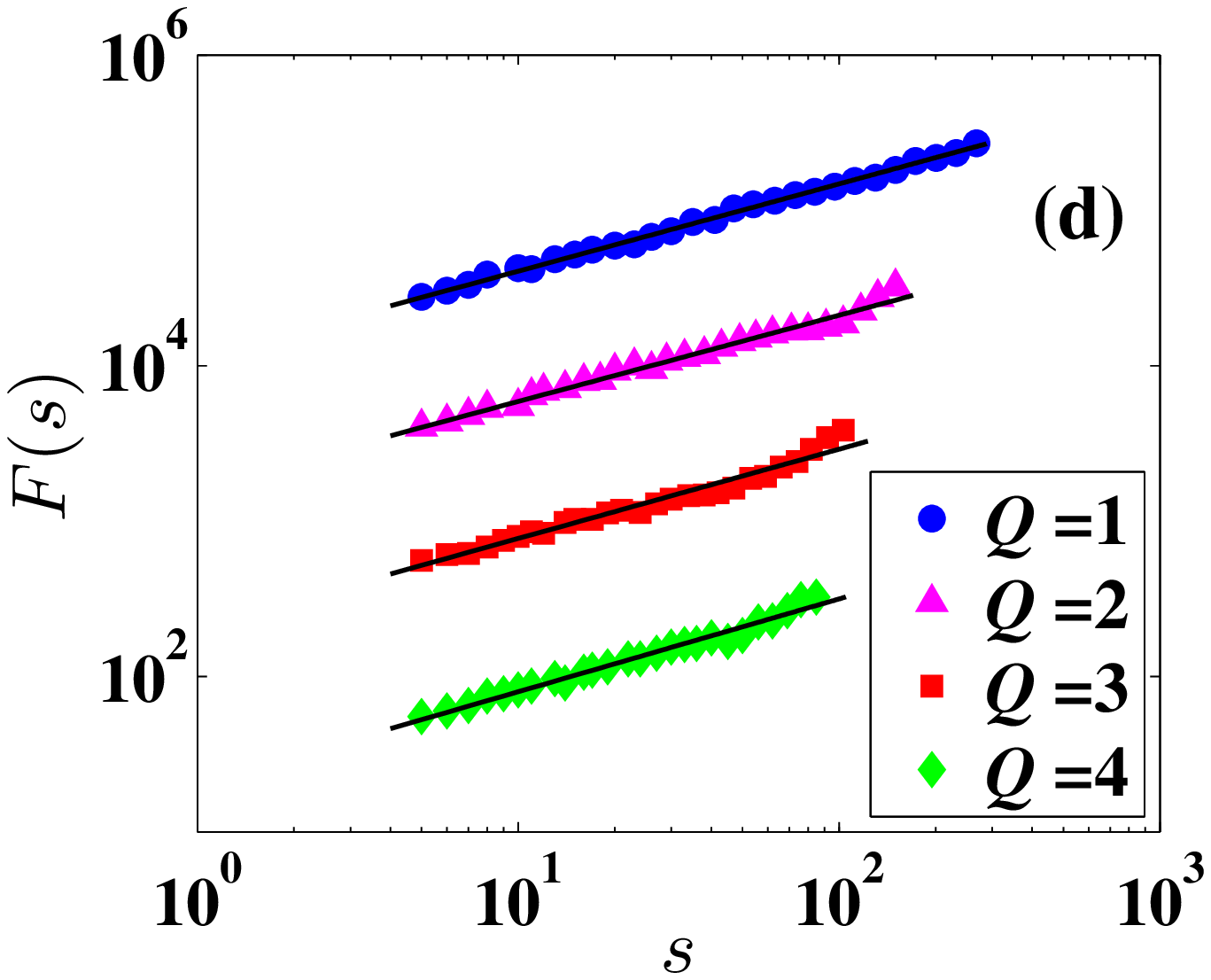}
\caption{\label{Fig:RI:MMF:DFA:Fs} (Color online) Detrended fluctuation functions $F(s)$ the return recurrence intervals above positive thresholds $Q=1,2,3$ and $4$ for four pairs of $H_s$ and $H_x$: (a) $H_x=0.5$ and $H_s=0.5$, (b)$H_x=0.75$ and $H_s=0.5$, (c) $H_x=0.5$ and $H_s=0.75$, and (d) $H_x=0.75$ and $H_s=0.75$. The curves have been vertically shifted for better visibility.}
\end{figure}

To investigate the long-term correlation in the recurrence intervals, we adopt the detrended fluctuation analysis (DFA) method
\cite{Peng-Buldyrev-Havlin-Simons-Stanley-Goldberger-1994-PRE,Kantelhardt-KoscielnyBunde-Rego-Havlin-Bunde-2001-PA,Hu-Ivanov-Chen-Carpena-Stanley-2001-PRE,Chen-Ivanov-Hu-Stanley-2002-PRE,Chen-Hu-Carpena-Bernaola-Galvan-Stanley-Ivanov-2005-PRE}. The DFA method calculates the detrended fluctuation function $F(s)$ of the time series within a window of $s$ points after removing a linear trend. For long-term power-law correlated time series, $F(s)$ is expected to scale as a power law with respect to the time scale $s$
\begin{equation}
   F(s)\sim s ^ H.
   \label{Eq:DFA}
\end{equation}
For each group of the input parameters, there are five series of return intervals for each $Q$. We calculate their $H_{Q,i}$ ($i=1,2,3,4,5$) separately and average them
\begin{equation}
  H_Q=\frac{1}{5}\sum_{i=1}^{5}H_{Q,i}.
\end{equation}

Figure \ref{Fig:RI:MMF:DFA:Fs} plots in double logarithmic scales the detrended fluctuation functions $F(s)$ of the return recurrence intervals above positive thresholds $Q=2$, $3$, $4$ and $5$ with respect to the scale $s$ for four typical pairs of $H_s$ and $H_x$. Power-law relations are observed and the associated DFA exponents $H_Q$ can be determined by linear regressions.

\begin{figure}[tb]
  \centering
  \includegraphics[width=4cm]{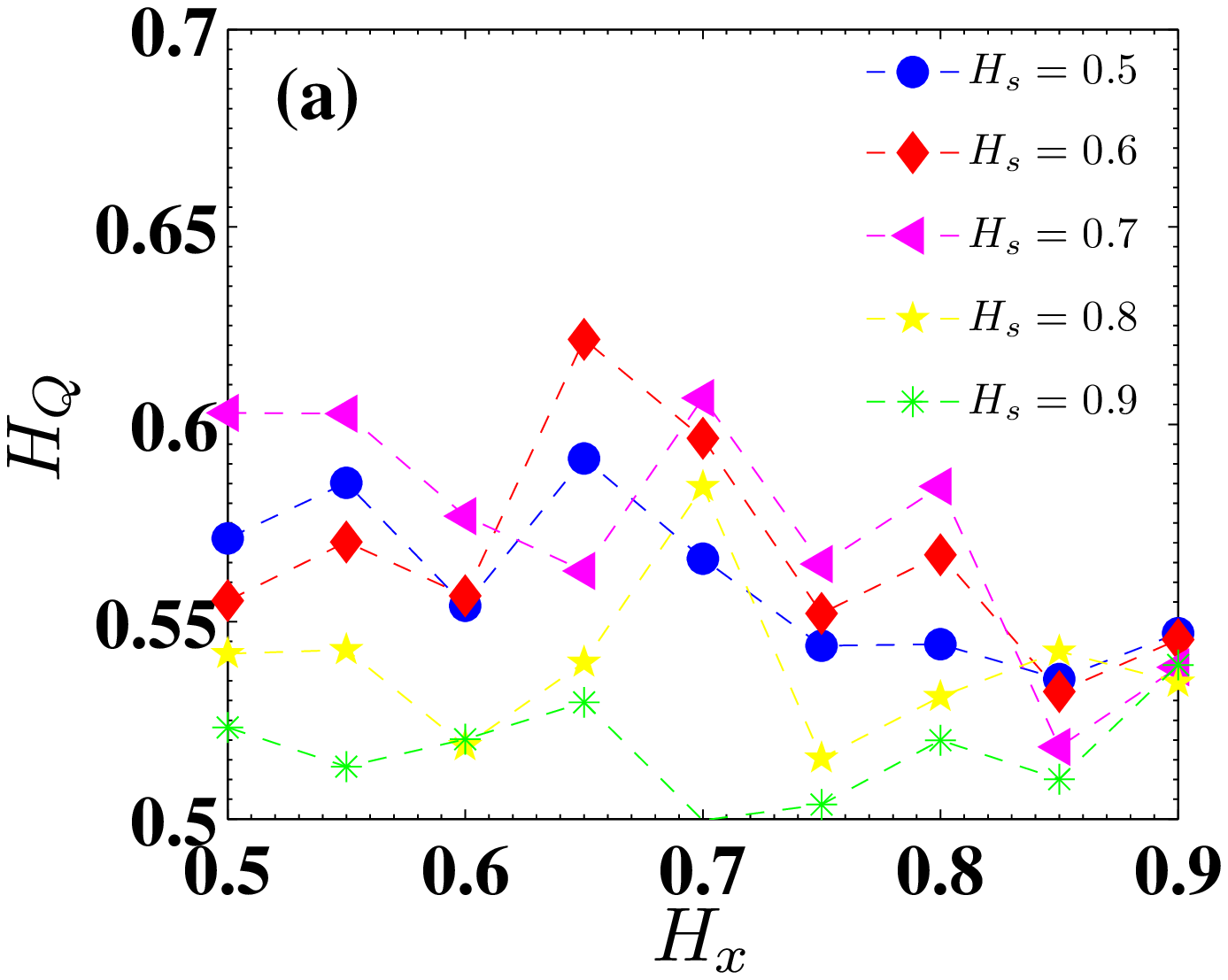}
  \includegraphics[width=4cm]{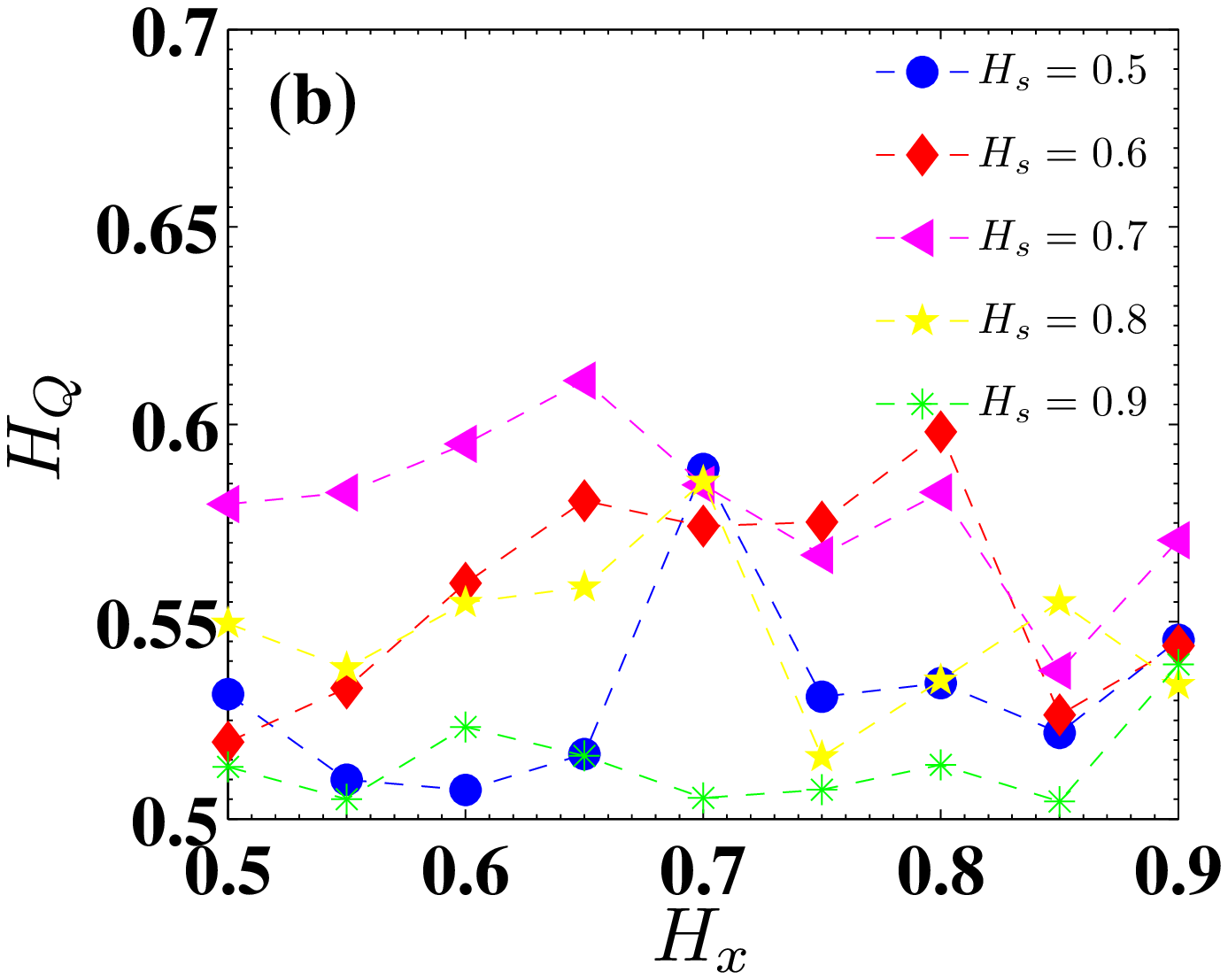}
  \includegraphics[width=4cm]{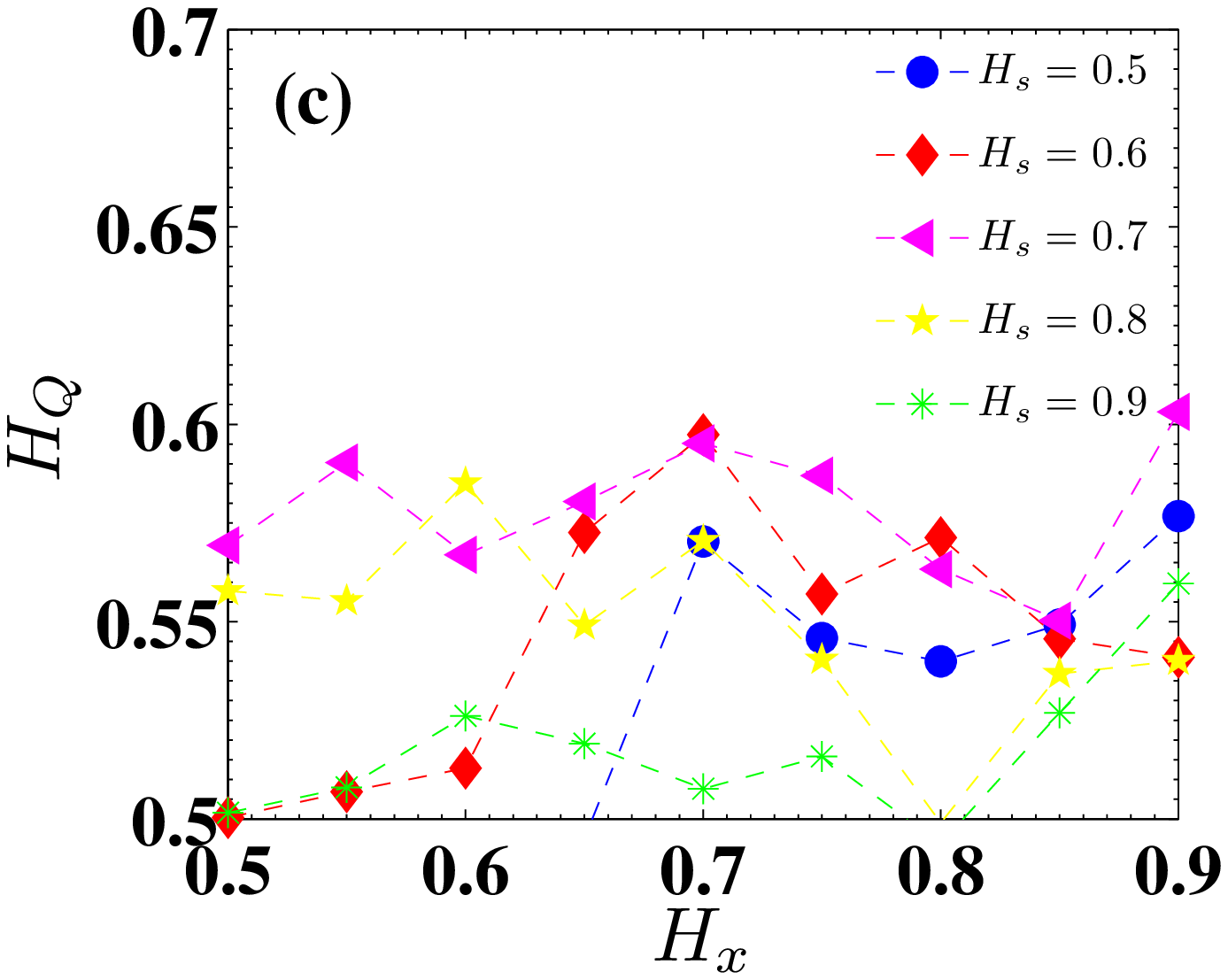}
  \includegraphics[width=4cm]{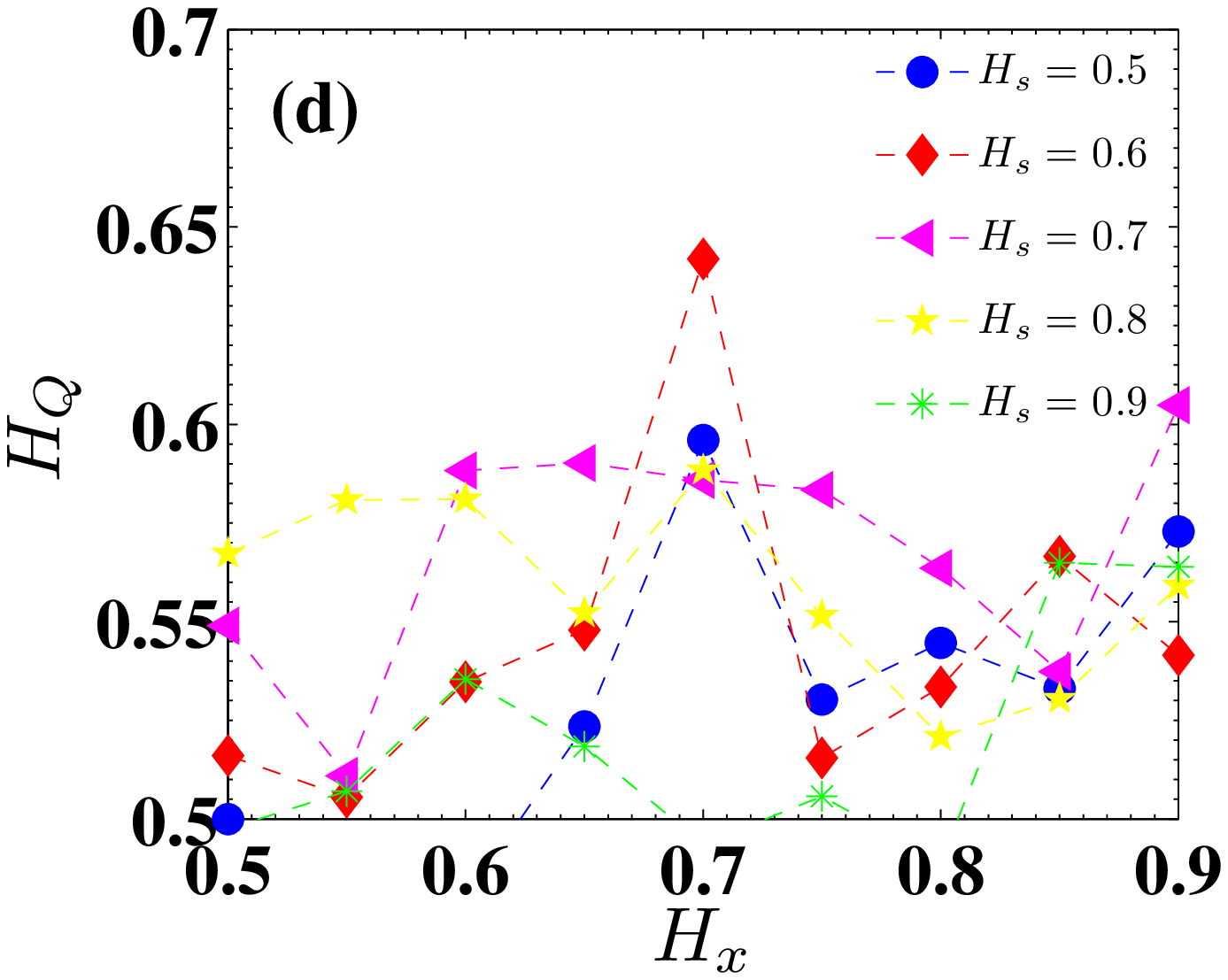}
  \caption{\label{Fig:RI:MMF:DFA:Hr:Hx} (Color online) Variation of the estimated DFA exponents $H_Q$ of recurrence interval series above different thresholds $Q=1$ (a), $Q=2$ (b), $Q=3$ (c) and $Q=4$ (d) with respect to $H_x$ for different $H_s$.}
\end{figure}

Figure \ref{Fig:RI:MMF:DFA:Hr:Hx} shows the dependence of the estimated DFA exponents $H_Q$ of recurrence interval series above different thresholds with respect to $H_x$ for different $H_s$. We find that the recurrence intervals exhibit weak memory effects since the $H_Q$ values are only slightly larger than 0.5. A careful scrutiny unveils that, for fixed $H_x$, $H_Q$ reaches a maximum when $H_s$ is close to 0.7. The DFA exponents $H_Q$ for the recurrence intervals of the shuffled return series fluctuate around 0.5 and the large $H_Q$ values are found to be significantly larger than 0.5. The observation of weak memory effects in the simulated recurrence intervals is inconsistent with the empirical finding \cite{Ren-Zhou-2010-NJP}, which indicates that certain important ingredients are missing in the modified Mike-Farmer model.

\section{Multifractality in recurrence intervals}
\label{S1:MF}

\begin{figure}
\centering
\includegraphics[width=4.3cm]{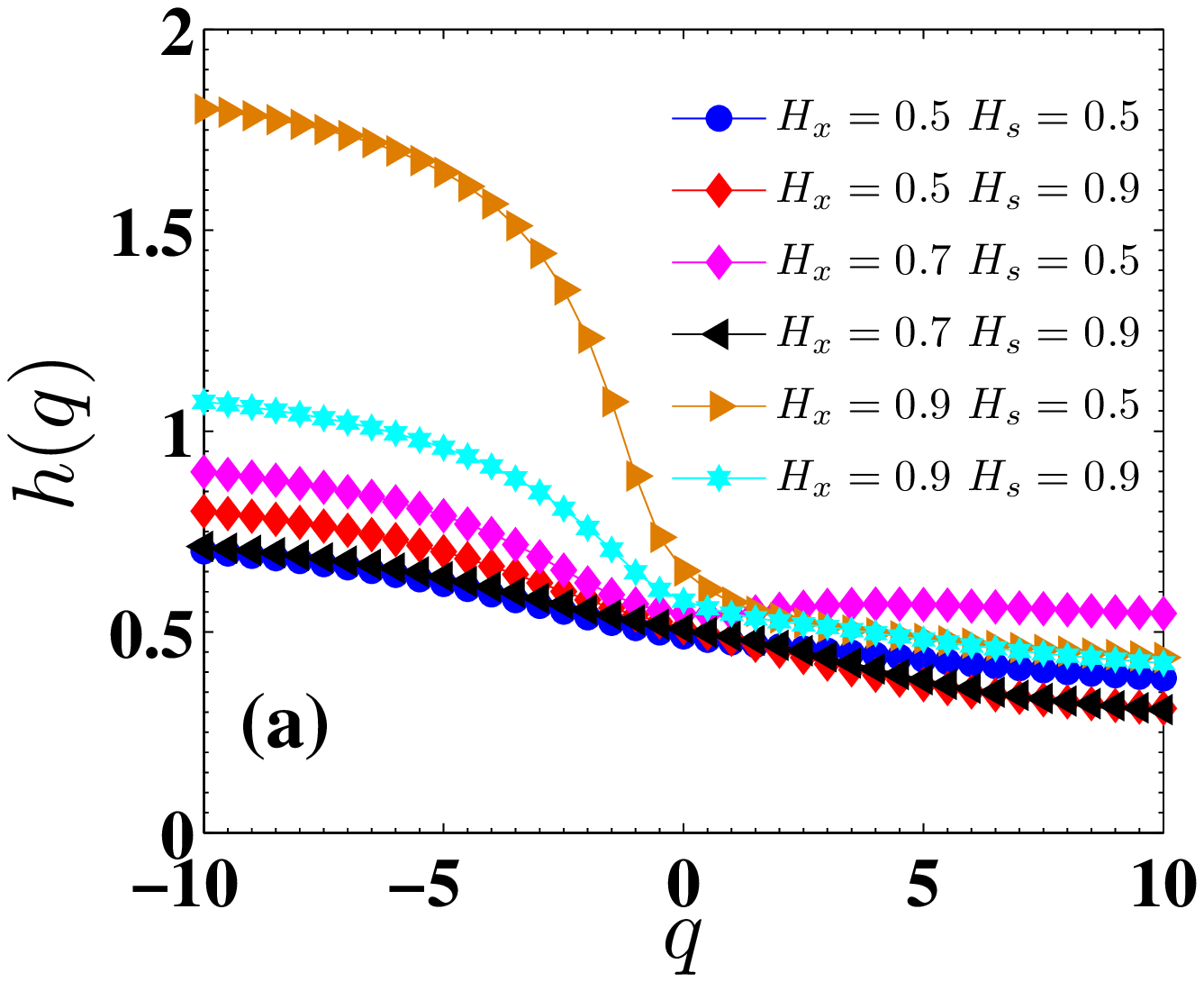}
\includegraphics[width=4.3cm]{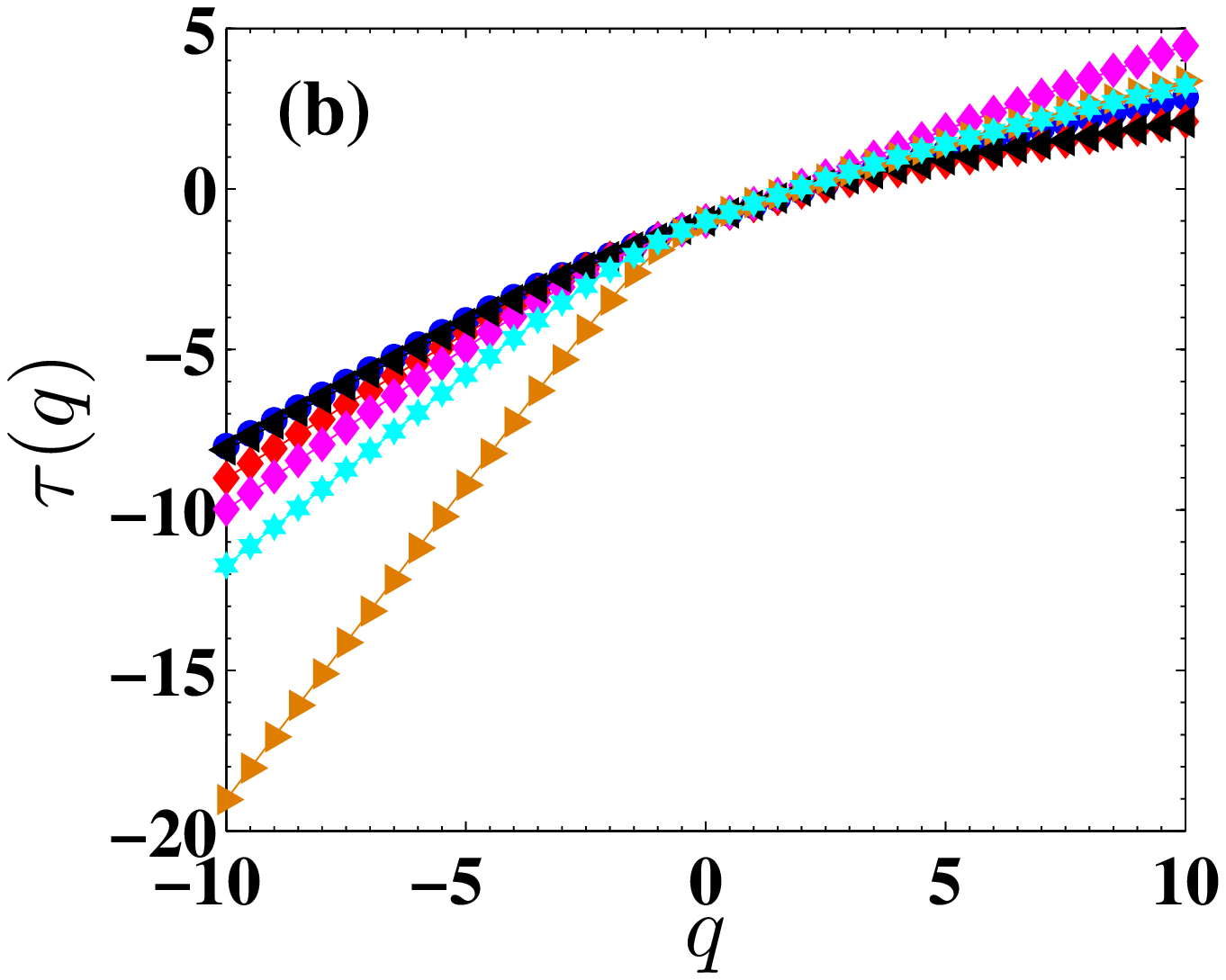}
\includegraphics[width=4.3cm]{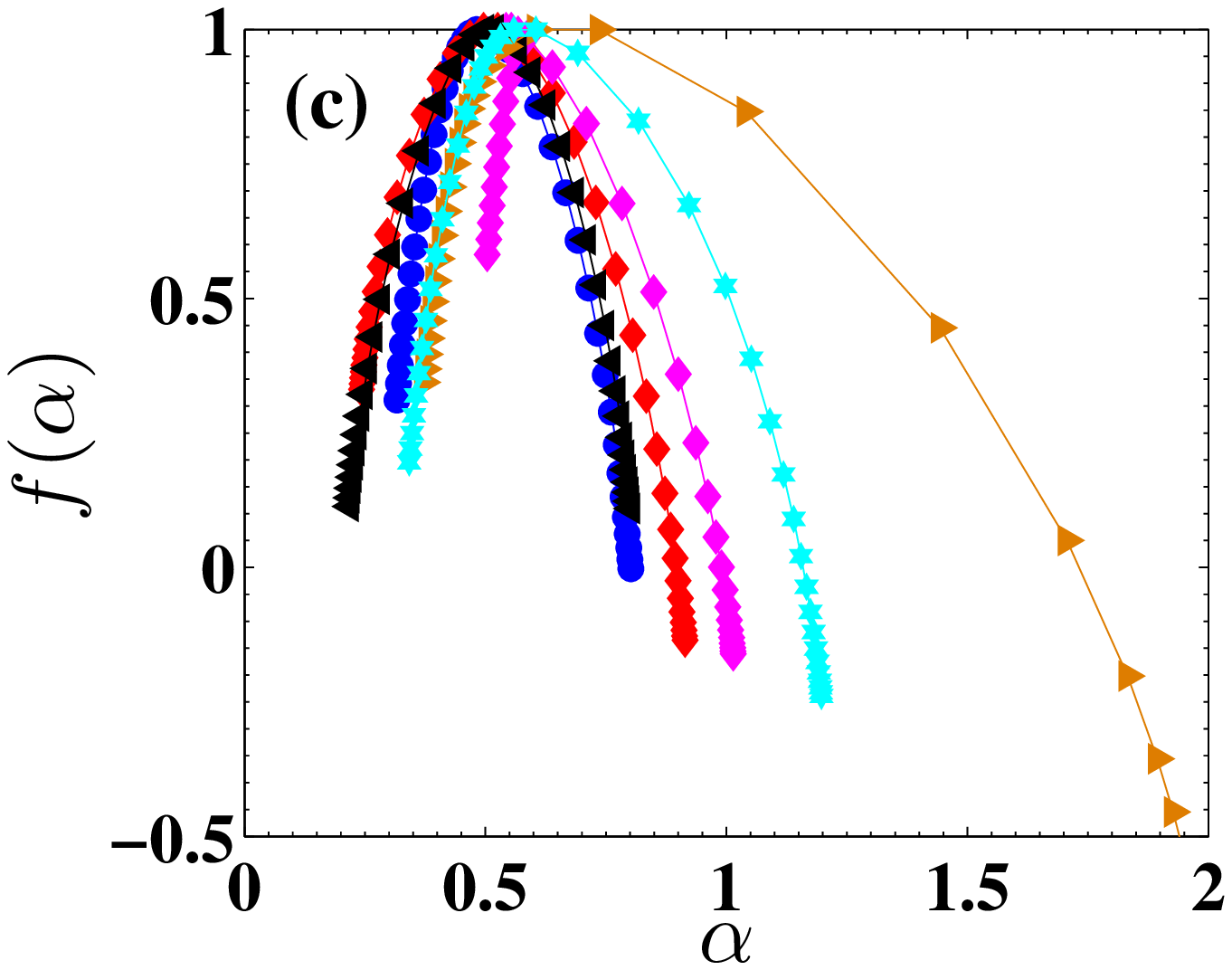}
\includegraphics[width=4.3cm]{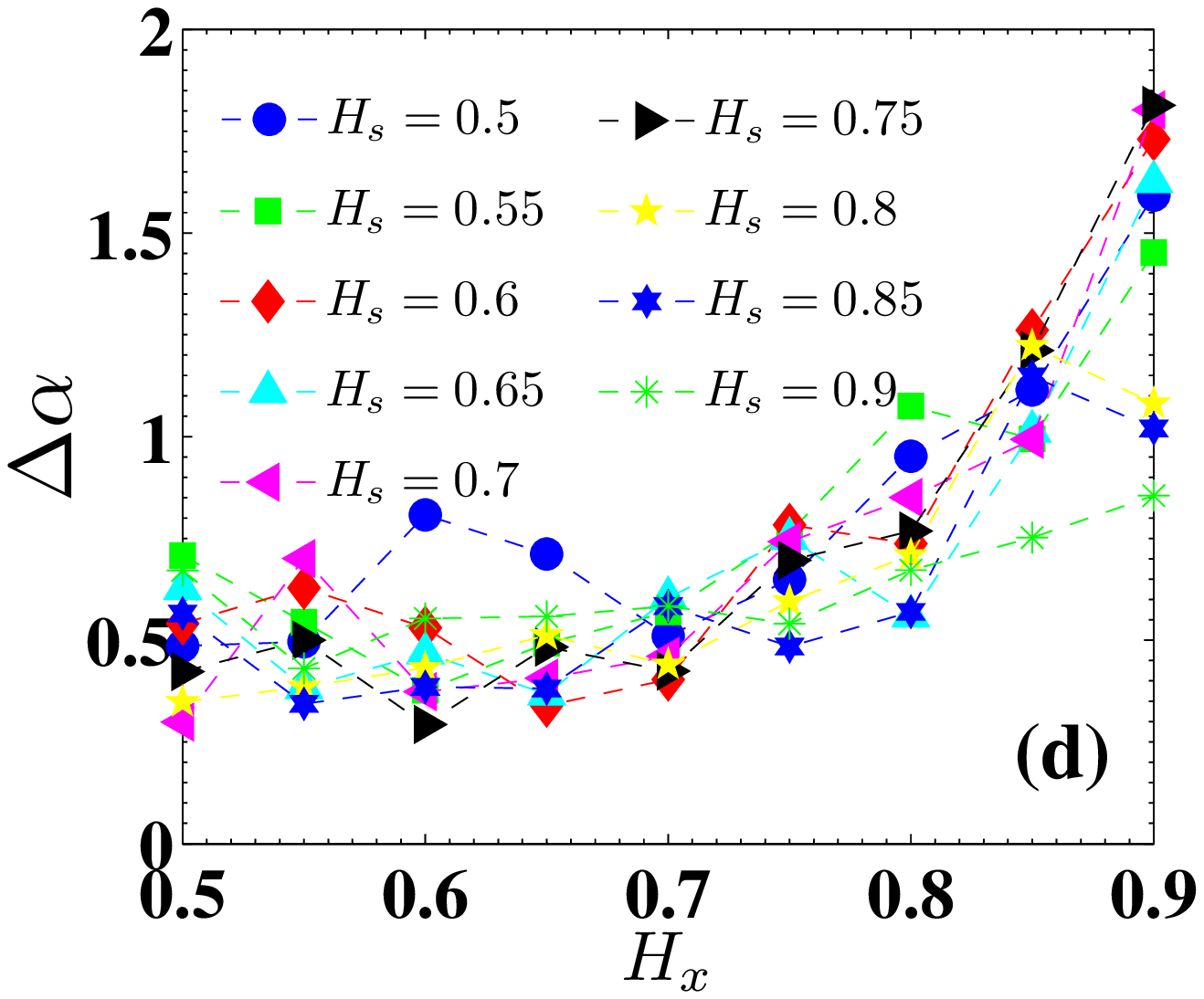}
\caption{\label{Fig:RI:MMF:MFDFA} (Color online) Multifractal analysis of recurrence intervals for $Q=4$. (a) MFDFA exponents $h(q)$ for typical pairs of $H_x$ and $H_s$. (b) Scaling exponents $\tau(q)$ calculated from the MFDFA exponents in plot (a). (c) Singularity spectrum $f(\alpha)$  calculated from the MFDFA exponents in plot (a). (d) Dependence of singularity width $\Delta\alpha$ on $H_x$ and $H_s$.}
\end{figure}

To our knowledge, the multifractal nature of recurrence intervals has only been investigated for energy dissipation rates \cite{Liu-Jiang-Ren-Zhou-2009-PRE}. Here, we adopt the multifractal detrended fluctuation analysis approach \cite{Kantelhardt-Zschiegner-KoscielnyBunde-Havlin-Bunde-Stanley-2002-PA} to investigate the multifractal nature of the simulated interval series. We expect that the $q$th order fluctuation function $F_{q}(s)$ scales with respect to the box size $s$ as a power law
\begin{equation}
  F_{q}(s) \sim s^{h(q)}.
\end{equation}
We can calculate the scaling exponents $\tau(q)$ as follows
\begin{equation}
 \tau(q) = q h(q) - 1.
\end{equation}
If $h(q)$ is a nonlinear function of $q$ (thus $\tau(q)$ is also a nonlinear function of $q$), the recurrence intervals exhibit multifractal nature. We can also calculate the singularity spectra by
\begin{equation}
  f(\alpha) = q [\alpha-h(q)]+1,
\end{equation}
where $\alpha = h(q)+q h^{\prime}(q)$ is the singularity strength. A broad spectrum of singularity indicates the presence of multifractality. Plots (a), (b) and (c) of Fig.~\ref{Fig:RI:MMF:MFDFA} illustrate respectively the $h(q)$, $\tau(q)$ and $f(\alpha)$ functions for several pairs of $H_x$ and $H_s$ for the threshold $Q=4$, in which each curve is averaged over five round of simulations. All these plots show that the recurrence interval series exhibit solid multifractal nature. According to Fig.~\ref{Fig:RI:MMF:MFDFA}(c), there are also negative $f(\alpha)$ values, known as negative or latent dimensions, which stem from random multifractals or lower-dimensional cuts of multifractal objects \cite{Mandelbrot-1990a-PA,Zhou-Yu-2001-PA,Zhou-Liu-Yu-2001-Fractals}.

In plots (a) to (c) of Fig.~\ref{Fig:RI:MMF:MFDFA}, we also notice that the curves differ from one another, especially for large values of $H_x$, which indicates that the multifractal nature of the associated recurrence intervals have different strength. We adopt the width of the singularity spectrum $\Delta\alpha$ to quantify the multifractal strength \cite{Zhou-2009-EPL}. The larger the singularity width, the stronger the multifractality. Figure~\ref{Fig:RI:MMF:MFDFA}(d) illustrates the dependence of  $\Delta\alpha$ with respect to $H_x$ for different $H_s$. No clear dependence is found between $\Delta\alpha$ and $H_s$. On the contrary, $\Delta\alpha$ remains stable when $0.5\leq H <0.7$ and increases for larger $H_x$. The results are qualitatively the same for other values of $Q$.

\section{Conclusions}

In summary, we have investigated the effects of long memory in the order submission process on the statistical properties of recurrence intervals of large price fluctuations through numerical experiments based on an order-driven stock trading model \cite{Gu-Zhou-2009-EPL}.

We found that the distributions of the scaled recurrence intervals of simulated returns can be well fitted with the generalized Gamma distribution, a power law with a stretched exponential decay. The Hurst index $H_x$ of the relative prices of submitted orders is found to have a significant impact on the power-law exponent $\beta$, while the Hurst index $H_s$ of order directions does not. Specifically, the power-law exponent $\beta$ increases linearly with the Hurst index $H_x$.

We also found that the recurrence intervals have weak long-term correlations based on the detrended fluctuation analysis, which is only partially consistent with empirical findings \cite{Ren-Zhou-2010-NJP}. No clear effects of $H_x$ and $H_s$ were observed on the Hurst index (or more precisely the DFA exponent) $H_Q$ of the recurrence intervals.

We further confirmed that the intervals possess multifractal nature. It is found that the long memory in the order directions has a negligible effect on the multifractal nature of recurrence intervals. In contrast, the singularity width of the multifractal nature fluctuates around a constant value when $H_x<0.7$ and then increases for larger values of $H_x$.

These findings indicate that the nontrivial properties of the recurrence intervals of returns are mainly caused by traders' behaviors of persistently placing new orders around the best bid and ask prices. It also implies that the original Mike-Farmer model \cite{Mike-Farmer-2008-JEDC} is not capable of investigating the topic in this Letter since it does not contain the ingredient of long memory in relative prices of submitted orders. We stress that the observation of weak memory effect in the simulated recurrence intervals calls for further improvements of the modified Mike-Farmer model \cite{Gu-Zhou-2009-EPL}.

\acknowledgments

This work was partially supported by the National Natural Science Foundation of China (Nos. 10905023, 11075054 and 71131007), the Humanities and Social Sciences Fund sponsored by Ministry of Education of the People's Republic of China (No. 09YJCZH042), the Program for Changjiang Scholars and Innovative Research Team in University (IRT1028), the Zhejiang Provincial Natural Science Foundation of China (No. Z6090130), and the Fundamental Research Funds for the Central Universities.

\bibliography{E:/Papers/Auxiliary/Bibliography}

\end{document}